\documentclass[aps,showpacs,preprintnumbers,amsmath,amssymb,eqsecnum,twocolumn,tightenlines]{revtex4}

\usepackage{graphicx}

\newcommand{\be}{\begin{eqnarray}}
\newcommand{\ee}{\end{eqnarray}}

 \newcommand{\gsim}{\mathrel{\hbox{\rlap{\lower.55ex \hbox {$\sim$}}
                   \kern-.3em \raise.4ex \hbox{$>$}}}}
\newcommand{\lsim}{\mathrel{\hbox{\rlap{\lower.55ex \hbox {$\sim$}}
                   \kern-.3em \raise.4ex \hbox{$<$}}}}

\def\roughly#1{\mathrel{\raise.3ex\hbox{$#1$\kern-.75em%
\lower1ex\hbox{$\sim$}}}}
\def\lsim{\roughly<}
\def\gsim{\roughly>}

\newcommand{\ba}{\begin{eqnarray}}
\newcommand{\ea}{\end{eqnarray}}

\begin{document}

\title{ Self-force and synchrotron radiation in odd space-time dimensions  }
\author { Edward Shuryak, Ho-Ung Yee, and Ismail Zahed }
\address { Department of Physics and Astronomy, State University of New York,
Stony Brook, NY 11794}
\date{\today}

\begin{abstract}
Classical electrodynamics in flat 3+1 space-time has a very special retarded propagator $\sim \delta(x^2)$ localized on the
light cone, so that a particle does not interact with its past field. However, this is an exception,
and in flat  odd-dimensional space-times as well as generic curved spaces this is not so. In this work
we show that the so called self-force
is not only non-zero, but  it matches  (in 2+1 dimensions) the radiation reaction force derived from
the radiation intensity. 
\end{abstract}
\maketitle

\section{Introduction}

This paper deals with 
 a century-old issue, the so called ``radiation reaction" force, in classical
electrodynamics and in general relativity. On one hand, it is clear that energy and momentum
carried away by radiation from an accelerated charge should be compensated by a force
which is going to reduce the energy and momentum of the particle accordingly.
On the other hand, the particle in 3+1  flat space-time does not interact with its own
field because it is fully concentrated on the light cone.  

The relation to the energy loss in the nonrelativistic dipole radiation 
and its relativistic extension  leads to the  well known Abraham-Lorentz-Dirac force
\be
f^\mu_{4D} = {2e^2 \over 3}\left(\dddot{x}^\mu - \ddot{x}^\nu \ddot{x}_\nu \dot{x}^\mu\right) \quad,\label{4d}
\ee
where the dot is a derivative over the proper time ${d/d\tau}$.  Its derivation comes from  ``the large-distance" discussion,  based on the amount of the energy/momentum fluxes through some distant surface
 (large sphere, etc). Although in principle such an
  approach can/was applied for scalar/electromagnetic/gravitational radiations, 
  in some cases it is technically difficult.  In particular, the radiation and its corresponding 
  ultrarelativistic  sources move through nearby paths in curved spaces, making the radiation
  calculation highly non-trivial (see e.g. \cite{Khriplovich:1973qf}) . It would be
   more satisfactory logically and much easier  practically to use some local derivation.
   In this paper we provide examples in support of a local derivation.

We stumbled on this issue while trying to assess the braking force for a gravitational
radiation of an ultrarelativistic particle moving
in a particular curved space (the so called thermal or black-hole AdS$_5$ in 4+1 dimensions).
This problem is related to the practically important problem of jet quenching in the quark-gluon plasma. 
Since the calculations are very different in AdS$_5$ we relegate the analysis
of jet quenching by braking radiation to the companion paper~\cite{Shuryak:2011ge}.

 The issue we study in this paper is whether one can define and calculate
 {\it the self-interaction force}, using only the particle's own field related to its past trajectory.
 This question was first addressed by Dirac for electrodynamics in flat 3+1 dimensions
 \cite{DIRAC} and extended to curved 3+1 dimensions by Dewitt~\cite{DEWITT}. In this work
 we show that this is also possible for electrodynamics in odd dimensional space-times,
 namely 2+1 and 4+1 dimensions.  
 
 The idea that it is possible was inspired by the  so called MSTQW approach  
 in gravitational setting \cite{Mino:1996nk,Quinn:1996am}. In it the
  remarkable  ``self-force" expression was suggested 
 \ba ma^a &=& m^2 u^b u^c \int_{-\infty}^{\tau^-}d\tau'  u^{a'} u^{b'} ( {1\over 2} \nabla^a G_{bca'b'} \\ \nonumber
 && - \nabla_b G_{c\,\,a'b'}^{\,\,a} -{1 \over 2} u^a u^d \nabla_d G_{bca'b'},
 ) \ea 
 in which the integral is done over proper time and the past  world line of the particle till the
 regulated present time $\tau^-$. $G$ is the retarded Green function for the Einstein equation
 with the particle as the source. Note that the bracket is just the Chrystoffel force for a gravity perturbation, 
 induced by the past gravity field of the particle itself. 
 
 Does this expression (or its simpler analogs) actually work? The first obvious try, for 
 electromagnetic radiation in 3+1 flat space-time, produces 
 zero because in flat 4d space-time the retarded propagator is
 totally localized on the light cone $(x^\alpha-r^\alpha)(x_\alpha-r_\alpha)=0$.  There are simply no points 
 on the particle path that can intersect the past light cone sustained by its present location! 
(It produces the well-known Lienard-Wiechert  expression for the retarded fields we all learned/taught in classical E/M courses).
This  upset can be remedied by realizing that such form of the retarded propagator is not the generic case,
and in fact  it is nonzero inside the light cone  in other space-times.
 Therefore, we decided to calculate it for the nearby space-time dimensions and see
 if the results make sense. We also calculate the radiative losses by standard large-distance method for comparison.
 
We assume that the motion of the charge is  prefixed by some external non-electromagnetic forces
and do not discuss its origin. We will  define and compute the local back-reaction force on a particle originating from its own electromagnetic fields. (Of course, this force should be a part of
the total force that determines the given charge trajectory, but we do not specify its effect on the path ). 

Although the self-force  is  in general expected to be nonlocal in character,
and given by the integral over the past trajectory, the situation is simplified in  
the highly relativistic limit, $\gamma\to\infty$.
Indeed we will see that in this case it  is defined by  a small range in the proper time interval $\tau\sim {1\over\gamma^2}$,
so that the leading $\gamma$-contribution depends only on the local data of the motion.
The resulting leading-$\gamma$ behavior of the self-force takes a similar form as in (\ref{4d}) 
in terms of local derivatives of the motion, which would be the odd-dimensional analogue of the
Abraham-Lorentz-Dirac force in 3+1 dimensions. 
Another feature is that the integrals with the retarded propagator diverge
and need to be correctly regulated: we will see this feature explicitly in our analysis shortly.

\section{The self-force in 2+1 dimensions \label{sec21}}

The case of 2+1 dimensions is the simplest one to consider. (Below we will see  
that it contains all the basic features of the radiation reaction force
in  any odd space-time dimensions because of some recursive relations between the propagators
involved.).
As already mentioned in the introduction, the main new feature in odd spacetimes compared to the conventional flat 3+1 dimensional space-time is that the massless retarded propagator from a given source has support $inside$  the light-cone. This implies that at a given moment/position of the charge, the electromagnetic
field acting on it obtain some contributions from  the past trajectory of the charge.

A moving charge $e$ with a given trajectory $x^\mu(\tau)$ gives a relativistic current
\be
j^\mu(x)=e\int d\tau\, \delta^{(3)}\left(x-x(\tau)\right) \dot{x}^\mu(\tau)\quad,\label{current}
\ee
where $\tau$ is the proper time normalized as $\dot{x}^\mu \dot{x}_\mu =+1$.
In the covariant gauge $\partial_\mu A^\mu=0$, the Maxwell equation becomes
\be
\Box A^\mu= j^\mu\quad,\quad \Box\equiv\partial_\mu\partial^\mu\quad,
\ee
which has a formal retarded solution as
\be
A^\mu(x)&=&\int d^3x'\, \Delta_R(x-x')j^\mu(x')\nonumber\\
&=&e\int d\tau'\,\Delta_R\left(x-x(\tau')\right)\dot{x}^\mu(\tau')\quad,\label{ret}
\ee
using (\ref{current}). $\Delta_R(x)$ is the massless retarded propagator in 2+1 dimensions which is given by
\be
\Delta_R(x)={\theta(x^0)\over 2\pi}{\theta(x^2)\over\sqrt{x^2}}\quad,\quad x^2\equiv x^\mu x_\mu\quad,\label{prop}
\ee
which has support in the entire forward light-cone.
It is clear that only the past trajectory contributes to the electromagnetic field at a given moment, and
we are interested in the self-force acting on the moving charge itself.
This amounts to computing the covariant Lorentz force 
\be
f^\mu_{3D}=e F^{\mu\nu}\left(x(\tau)\right) \dot{x}_\nu(\tau) \quad,\label{3d}
\ee
where $F^{\mu\nu}\left(x(\tau)\right)$ are the field strengths of the retarded electromagnetic field (\ref{ret}) 
at proper time $\tau$, induced by
the past trajectory of the charge, $x^\mu(\tau')$ with $\tau'<\tau$.

As usual, one encounters local divergences in computing (\ref{3d}) coming from the region near $\tau'=\tau$, and one has to regularize
and absorb divergences by renormalizing physical parameters of the moving charge such as its mass.
In this section, we regularize divergences covariantly by cutting off proper time integral 
\be
\int^\tau_{-\infty} d\tau' \to \int^{\tau-\epsilon}_{-\infty}d\tau'\quad,\label{cutoff}
\ee
with a small $\epsilon>0$, and let $\epsilon\to 0$ after renormalizing the particle mass.
In the appendix, we will present another regularization by averaging (\ref{3d}) over a small sphere
around $x(\tau)$ of radius $r=\epsilon$ and taking $\epsilon\to 0$ after renormalization. The results are identical, which
is at least a useful consistency check of our results. 

From (\ref{ret}), (\ref{3d}), and using the fact that
\be
\partial^\mu \Delta_R(x)=2 x^\mu \Delta_R'(x)+\delta^{\mu 0}{\delta(x^0)\over 2\pi}{\theta(x^2)\over\sqrt{x^2}}\quad,
\ee
where 
\be
\Delta_R'(x)=\frac{\theta(x^0)}{2\pi}\left(\frac{\delta(x^2)}{\sqrt{x^2}}-\frac 12\,\frac{\theta(x^2)}{x^2\sqrt{x^2}}\right)\quad,
\label{T6}
\ee
the unregularized bare Lorentz force is
\be
&&f^\mu_{3D}=2e^2\int d\tau'\,\Delta_R'\left({\bf X}\right) {\bf X}^{[\mu}(\tau,\tau')\dot{x}^{\nu]}(\tau')\dot{x}_\nu(\tau)\nonumber\\
&&+ e^2\delta^{[\mu 0}\int d\tau'\,{\delta\left({\bf X}^0\right)\over 2\pi}{\theta\left({\bf X}^2\right)\over \sqrt{{\bf X}^2}} \dot{x}^{\nu]}(\tau')\dot{x}_\nu (\tau)\quad,\label{unreg}
\ee
where ${\bf X}^\mu(\tau,\tau')\equiv x^\mu(\tau)-x^\mu(\tau')$ and $[\mu,\nu]=\mu\nu-\nu\mu$.
To identify the local divergences near $\tau'=\tau$, we expand the quantities in terms of $(\tau-\tau')\equiv\epsilon$ as
\be
&&{\bf X}^\mu(\tau,\tau')=\epsilon\dot{x}^\mu(\tau) -{\epsilon^2\over 2}\ddot{x}^\mu(\tau)+{\epsilon^3\over 6}\dddot{x}^\mu(\tau)+{\cal O}(\epsilon^4)\,,\nonumber\\
&&\dot{x}^\nu(\tau')=\dot{x}^\nu(\tau)-\epsilon\ddot{x}^\nu(\tau)+{\epsilon^2\over 2}\dddot{x}^\nu(\tau)+{\cal O}(\epsilon^3)\,,
\nonumber\\
&& {\bf X}^2=\epsilon^2-{\epsilon^4\over 12}\ddot{x}^\mu\ddot{x}_\mu +{\cal O}(\epsilon^5)\,,\label{epsexp}
\ee
using the identities such as $\dot{x}^\mu\ddot{x}_\mu=0$ and $\dot{x}^\mu\dddot{x}_\mu=-\ddot{x}^\mu\ddot{x}_\mu$.
From these, one obtains after some algebra
\be
&&{\bf X}^{[\mu}(\tau,\tau')\dot{x}^{\nu]}(\tau')\dot{x}_\nu(\tau)\\\label{expand}
&&={\epsilon^2\over 2}\ddot{x}^\mu(\tau)-{\epsilon^3\over 3}\left(\dddot{x}^\mu+\ddot{x}^\nu\ddot{x}_\nu\dot{x}^\mu\right)(\tau) +{\cal O}(\epsilon^4)\quad,\nonumber
\ee
and
\be
\Delta_R'({\bf X})=-{\theta(\epsilon)\over 4\pi}\left({1\over\epsilon^3}-{\delta(\epsilon)\over \epsilon^2}\right) +{\cal O}\left(1\over\epsilon\right)\quad,
\ee
so that the proper time integral of the first term in (\ref{unreg}) near the $\epsilon=0$ region becomes
\be
-{e^2\over 4\pi}\int d\epsilon\,\theta(\epsilon)\left({1\over\epsilon}-\delta(\epsilon)\right)\ddot{x}^\mu(\tau) +{\cal O}(\epsilon^1)\quad,\label{loginf}
\ee
which is logarithmically divergent.
This divergence is proportional to the covariant acceleration $\ddot{x}^\mu$ and is readily absorbed by renormalizing the
mass
\be
m_{ren}=m_{bare}-{e^2\over 4\pi}\log(\epsilon)\quad,\label{mren}
\ee
where $\epsilon$ here means a cutoff with a slight abuse of notation.
This is equivalent to having a counterterm added to (\ref{unreg}) so that the renormalized self-force now takes a form
\be
&&f^\mu_{3D}=-{e^2\over 2\pi}\lim_{\epsilon\to 0^+}\int d\tau'\,\theta(\tau-\tau'-\epsilon)\label{renormalized}\\&&\times\left({{\bf X}^{[\mu}(\tau,\tau')\dot{x}^{\nu]}(\tau')\dot{x}_\nu(\tau)
\over\left({\bf X}^2\right)^{3\over 2}} -{1\over 2}{\ddot{x}^\mu(\tau)\over (\tau-\tau')}\right)\quad,\nonumber
\ee
considering only the first term in (\ref{unreg}).

The finite contribution coming from the piece proportional to $\delta(\epsilon)$ in (\ref{loginf}) is ambiguous up to the value of $\theta(0)$.
In fact, the second term in (\ref{unreg}) gives rise to a divergence which is also proportional to $\theta(0)$.
One can choose to have $\theta(0)=0$ removing these ambiguities as a choice of regularization. Alternatively, once we introduce
a cutoff $\tau-\tau'>\epsilon$ they simply don't appear for any finite $\epsilon>0$, and hence the limit $\epsilon\to 0^+$
after taking care of the logarithmic divergence via (\ref{mren}) is insensitive to them.

Unlike in 3+1 dimensions where the self-force is the local
Abraham-Lorentz-Dirac force (\ref{4d}), in 2+1 dimensions the self-force (\ref{renormalized}) is non-local and receives contributions
from the entire past tail of the charged particle. This non-locality is due to the peculiar fact that the retarded propagator 
(\ref{prop}) has support on the entire forward part of the light cone as we noted earlier.

However, for ultra-relativistic motion we now show that the leading $\gamma$ behavior of (\ref{renormalized})
is completely local. Let's denote the $n$-th derivative of $x^\mu$ with respect to proper time $\tau$ as $x_{(n)}^\mu\equiv
{d^n x^\mu\over d\tau^n}$. In the large $\gamma$-limit, $x^\mu_{(n)}$ is of order ${\cal O}(\gamma^n)$.
In general, for any combinations of $x^\mu_{(n)}$ the powers of $\gamma$ simply add up
\be
\Pi_i x_{(n_i)} \le {\cal O}\left(\gamma^{\sum_i n_i}\right) \quad,\label{powers}
\ee
where we allow the inequality due to exceptions of some lowest contractions such as $\dot{x}^\mu\dot{x}_\mu=1$ and $\dot{x}^\mu \ddot{x}_\mu=0$.
These exceptions will not affect our derivation and conclusion, because what will be important  
is that the maximum powers of $\gamma$ is $\sum_i n_i$.
We will prove that the leading $\gamma$ behavior of (\ref{renormalized}) is coming from the proper time interval
$\tau-\tau'\equiv \epsilon\sim {1\over\gamma^2}$: we will first assume this and expand (\ref{renormalized})
in small $\epsilon$, then show the consistency of the assumption in our final results.

Using (\ref{epsexp}) and (\ref{powers}), one can show the expansion
\be
&&{\bf X}^{[\mu}(\tau,\tau')\dot{x}^{\nu]}(\tau')\dot{x}_\nu(\tau)\\\label{expand2}
&&={\epsilon^2\over 2}\ddot{x}^\mu(\tau)-{\epsilon^3\over 3}\left(\dddot{x}^\mu+\ddot{x}^\nu\ddot{x}_\nu\dot{x}^\mu\right)(\tau) +{\cal O}(\epsilon^4\gamma^6)\quad,\nonumber\\
&& {\bf X}^2=\epsilon^2-{\epsilon^4\over 12}\ddot{x}^\nu\ddot{x}_\nu +{\cal O}(\epsilon^5\gamma^5)\quad,
\ee
Inserting the expansion into (\ref{renormalized}), yields

\be
&&{1\over 2\epsilon}\left({1\over\left(1-{\epsilon^2\over 12}\ddot{x}^\nu\ddot{x}_\nu\right)^{3\over 2}}
-1\right)\ddot{x}^\mu\\&&
-{1\over 3\left(1-{\epsilon^2\over 12}\ddot{x}^\nu\ddot{x}_\nu\right)^{3\over 2}}\left(\dddot{x}^\mu+\ddot{x}^\nu\ddot{x}_\nu\dot{x}^\mu\right)\nonumber +{\cal O}\left(\gamma^4\right)\quad,
\ee
In estimating ${\cal O}\left(\gamma^4\right)$ we already used the assumption $\epsilon\sim {1\over\gamma^2}$.
This is justified because of the expression in the denominator which effectively cutoffs the $\epsilon$-integral 
and confines it to $\epsilon\sim (\ddot{x})^{-1}\sim {1\over\gamma^2}$.
The dominant contribution in the large $\gamma$ limit then comes from the second line which is ${\cal O}\left(\gamma^5\right)$,
while others are all ${\cal O}\left(\gamma^4\right)$, so that the leading self-force becomes
\be
f^{\mu}_{3D}&\approx&{e^2\over 6\pi}\left(\int_0^\infty{d\epsilon\over
\left(1-{\epsilon^2\over 12}\ddot{x}^\nu\ddot{x}_\nu\right)^{3\over 2}}\right) \ddot{x}^\nu\ddot{x}_\nu \dot{x}^\mu+{\cal O}(\gamma^2)\nonumber\\
&=& -{e^2\over\sqrt{3}\pi}\sqrt{-\ddot{x}^\nu\ddot{x}_\nu} \,\,\dot{x}^\mu+{\cal O}(\gamma^2)\quad,
\ee
which is ${\cal O}\left(\gamma^3\right)$. This is our 2+1 dimensional version of the Abraham-Lorentz-Dirac force.

We apply our result to the ultra-relativistic circular motion
\be
x^\mu(\tau)=\left(\gamma\tau, \rho\,{\rm cos}(\gamma\omega\tau), \rho\,{\rm sin}(\gamma\omega\tau)\right)\quad,
\label{T10}
\ee 
where ${1\over\gamma^2}=(1-v^2)\ll 1$ with $v=\rho\,\omega\approx 1$.
Because $\ddot{x}^\nu \ddot{x}_\nu=-\rho^2 \omega^4\gamma^4 \approx -\omega^2 \gamma^4$, the proper time integral is
confined to
\be
|\tau'-\tau|\approx\frac 1{\omega\gamma^2}\ll 1\quad,
\label{T11}
\ee 
as discussed before, and the leading covariant self-force is
\be
f^\mu_{3D}=-{e^2\over\sqrt{3}\pi} \omega\gamma^2 \dot{x}^\mu\label{ressec2}
\quad,
\ee
which is longitudinal.
The common
(non-covariant) longitudinal drag force is
\be
\vec{\bf f}_L\approx -\frac{e^2}{\sqrt{3}\pi}\omega\gamma^2\vec{v}\quad.
\label{T14}
\ee
Therefore, the external force that is needed to maintain the circular motion is
\be
\vec{\bf f}_{ext}= m_{ren}{d^2\vec{x}\over dt^2} -\vec{\bf f}_L\quad,
\ee
and the rate of work done to the particle is given by
\be
P=\vec{\bf f}_{ext}\cdot\vec{v}\approx \frac{e^2}{\sqrt{3}\pi}\omega\gamma^2\quad,
\label{work}
\ee
where we used ${d^2\vec{x}\over dt^2}\cdot\vec{v}=0$ for the circular motion.

In section (\ref{farsec}), we will compute the far-field radiation of this synchrotron motion, and find
an agreement between the total power radiated at large distance and the work done locally to the charge (\ref{work}).

\section{Local self-force in 4+1 dimensions}

In this section, we will perform similar computations as in the previous section but in 4+1 dimensional spacetime.
We will therefore be brief in explaining the details of the derivation while presenting our results.
The Maxwell's equations sourced by a charge $e$ in the radiative gauge are solved as before,
\be
A^\mu(x)
&=&e\int d\tau'\,\Delta_R\left(x-x(\tau')\right)\dot{x}^\mu(\tau'),\label{ret5d}
\ee
where the retarded propagator in 4+1 dimensions is
\be
\Delta_R(x)=
-\frac{\theta(x^0)}{2\pi^2}\left(\frac{\delta(x^2)}{\sqrt{x^2}}-\frac 12\,\frac{\theta(x^2)}{x^2\sqrt{x^2}}\right).
\label{F3}
\ee
Note the analogy with (\ref{prop}), which in fact stems from the generic relationship 
\be
\Delta_{4+1}(x)=-\frac 1\pi\,\frac{d}{dx^2}\,\Delta_{2+1}(x),
\label{F4}
\ee
which is readily shown from the momentum space representation of the retarded
propagators and the recursive property of the integer Bessel functions. Modulo 
normalizations, (\ref{F4}) extends to all odd space-time dimensions. 

From (\ref{ret5d}) and (\ref{F3}), one easily writes down the unregularized self-force simialr to (\ref{unreg}) before as  
\be
f^\mu_{5D}=
2e^2\int d\tau'\,\Delta_R'\left({\bf X}\right) {\bf X}^{[\mu}(\tau,\tau')\dot{x}^{\nu]}(\tau')\dot{x}_\nu(\tau),\label{unreg5d}
\ee
where in 4+1 dimensions we have 
\be
\Delta'_R(x)=-\frac{\theta(x^0)}{2\pi^2}\left(\frac{\delta'(x^2)}{\sqrt{x^2}}-\frac{\delta(x^2)}{(x^2)^{3\over 2}}
+\frac 34 \frac{\theta(x^2)}{(x^2)^{5\over 2}}\right).
\label{F6}
\ee
In (\ref{unreg5d}), we have dropped terms that are proportional to $\delta\left({\bf X}^0\right)$ due to our regularization scheme that we explain in the following. However, we should caution the readers that we haven't checked the regularization scheme independency
of our results, contrary to the previous 2+1 dimensional case.

We will regularize the divergences appearing in (\ref{unreg5d}) by replacing $\Delta'_R(x)$ with
\be
\Delta^{'\epsilon}_R(x)=-\frac{\theta(x^0)}{2\pi^2}\left(\frac{\delta'(x^2-\epsilon^2)}{\sqrt{x^2}}-\frac{\delta(x^2-\epsilon^2)}{(x^2)^{3\over 2}}
+\frac 34 \frac{\theta(x^2-\epsilon^2)}{(x^2)^{5\over2}}\right),\nonumber\\
\label{regprop}
\ee
and taking the $\epsilon\to 0$ limit after removing divergences by renormalization.
To identify the divergences in (\ref{unreg5d}), the necessary small $\epsilon$-expansion reads as
\be
&&{\bf X}^{[\mu}(\tau,\tau')\dot{x}^{\nu]}(\tau')\dot{x}_\nu(\tau)\label{expand5d}\\
&&={\epsilon^2\over 2}\ddot{x}^\mu(\tau)-{\epsilon^3\over 3}\left(\dddot{x}^\mu+\ddot{x}^\nu\ddot{x}_\nu\dot{x}^\mu\right)(\tau)
\nonumber\\
&& +{\epsilon^4\over 8}\left(\ddddot{x}^\mu-\ddddot{x}^\nu \dot{x}_\nu \dot{x}^\mu +{2\over 3} \ddot{x}^\nu\ddot{x}_\nu \ddot{x}^\mu\right)\nonumber\\
&&-{\epsilon^5\over 30}\left({x}^{(5)\mu}-{x}^{(5)\nu} \dot{x}_\nu \dot{x}^\mu -{5\over 4} \ddddot{x}^\nu\dot{x}_\nu \ddot{x}^\mu\right)
+{\cal O}(\epsilon^6\gamma^8),\nonumber
\ee
where $x^{(5)}\equiv {d^5 x\over d\tau^5}$.
Upon inserting (\ref{expand5d}) into (\ref{unreg5d}), the divergences of the self-force are found to be
\be
f^\mu_{5D}&\sim& {e^2\over 16\pi^2}{1\over\epsilon^2}\ddot{x}^\mu\label{leading}\\
&+&{3e^2\over 32\pi^2}\log(\epsilon)\left(\ddddot{x}^\mu-\ddddot{x}^\nu \dot{x}_\nu \dot{x}^\mu +{2\over 3} \ddot{x}^\nu\ddot{x}_\nu \ddot{x}^\mu\right).\nonumber
\ee
The first leading divergence can readily be absorbed into the renormalized mass
\be
m_{ren}=m_{bare}-{e^2\over 16\pi^2}{1\over\epsilon^2},
\ee
whereas the nature of the second term of logarithmic divergence is unclear to us. We will simply choose our regularization
scheme to remove it minimally. We expect this minimal subtraction to be consistent with the far-field radiation formulae,
although we will only check this for the 2+1 dimensional case below.

After removing the divergences, the finite contribution can be computed in the leading $\gamma$ approximation.
One finds that the $\delta({\bf X^2})$ and $\delta'({\bf X^2})$ terms in $\Delta'_R({\bf X})$ in (\ref{unreg5d})
give us the leading $\gamma^6$ contributions to the self-force, 
\be
f^\mu_{5D}\sim{e^2\over 8\pi^2}\left(\ddddot{x}^\mu+\ddot{x}^\nu\ddot{x}_\nu \ddot{x}^\mu- \ddddot{x}^\nu\dot{x}_\nu \dot{x}^\mu\right)+{\cal O}(\gamma^5),
\ee
which is completely local. In deriving the above, one has to expand
\be
{\bf X}^2=\epsilon^2-{\epsilon^4\over 12}\ddot{x}^\mu\ddot{x}_\mu 
-{\epsilon^5\over 12}\left(\ddddot{x}^\mu\dot{x}_\mu +2 \ddot{x}^\mu\dddot{x}_\mu\right)+{\cal O}(\epsilon^6\gamma^6)\nonumber.
\ee
However, for circular motion that we are interested in,  the longitudinal
component of the above force that is related to the rate of work done simply vanishes. Therefore, we are led to seek
the next leading term in ${\cal O}(\gamma^5)$.

The next leading term is quasi-local, that is, it comes from the region $\epsilon\sim{1\over\gamma^2}$ as was the case in 2+1 dimensions before.
It is given by
\be
&&-{e^2\over 40\pi^2}\left(\int_0^\infty d\epsilon\,{1\over\left(1-{\epsilon^2\over12}\ddot{x}^\nu\ddot{x}_\nu\right)^{5\over 2}}\right)\\&&\times\big(x^{(5)\nu}\dot{x}_\nu \dot{x}^\mu +{35\over 8}\ddddot{x}^\nu\dot{x}_\nu \ddot{x}^\mu +{25\over 4}\ddot{x}^\nu\dddot{x}_\nu\ddot{x}^\mu\big)\nonumber\\
&&=-{e^2\over 10\sqrt{3}\pi^2}{\big(x^{(5)\nu}\dot{x}_\nu \dot{x}^\mu +{35\over 8}\ddddot{x}^\nu\dot{x}_\nu \ddot{x}^\mu +{25\over 4}\ddot{x}^\nu\dddot{x}_\nu\ddot{x}^\mu\big)\over\sqrt{-\ddot{x}^\nu\ddot{x}_\nu}}.\nonumber
\ee 

Applying the above to the ultra-relativistic synchrotron motion, we see that only the first term contributes, and we obtain the leading longitudinal force as
\be
f^\mu_{5D}\sim -{e^2\over 10\sqrt{3}\pi^2}\gamma^4\omega^3 \dot{x}^\mu,
\ee
which leads to the rate of work done to the system,
\be
P={e^2\over 10\sqrt{3}\pi^2}\gamma^4\omega^3 .
\ee

\section{Far-field synchrotron radiation in 2+1 dimensions\label{farsec}}

In this section, we compute the far-field radiation of the ultra-relativistic circular motion
and check that the resulting total rate of radiation matches the rate of work done locally to the charge (\ref{work}),
which is our first nontrivial  consistency check for the ``self-force" approach and the subtractions we have applied.

The retarded gauge field from a point-like charge motion is
\be
A^\mu(t,\vec x)={e\over 2\pi}\int^{t_*(t,\vec x)}_{-\infty}dt'\,{1\over \sqrt{(t-t')^2- (\vec x-\vec x'(t'))^2}} {dx'^\mu\over dt'},\nonumber
\ee
where $t_*(t,\vec x)<0$ is determined by the intersection between the past light-cone from $(t,\vec x)$ and the particle trajectory
\be
(t-t_*)^2-(\vec x-\vec x'(t_*))^2=0\quad.\label{deft}
\ee
For the synchrotron motion we are interested in
\be
x^\mu&=&(t,\rho\cos(\omega t),\rho\sin(\omega t))\nonumber\\
&=&\left(\gamma\tau, \rho\,{\rm cos}(\gamma\omega\tau), \rho\,{\rm sin}(\gamma\omega\tau)\right)\quad,
\ee
the field strengths can be written explicitly as in the appendix. We will start our discussion with the formulae (\ref{app10}), (\ref{app11}), and (\ref{app12}).
We are interested in the far-field asymptotics at $r\to\infty$, where we introduce a polar coordinate system $(r,\phi)$ on the spatial two dimensional plane. The expression for the total far-field radiation is
\be
{d E\over dt}&=& \lim_{r\to\infty}r\int_0^{2\pi}d\phi\, T_{0i}\hat n^i\label{4s1}\\
&=&\lim_{r\to\infty}r\int_0^{2\pi}d\phi\, F_{12}\left(F_{01}\sin\phi-F_{02}\cos\phi\right),\nonumber
\ee
where $\hat n=(\cos\phi,\sin\phi)$ is the unit radial vector.

To compute the leading large $r$ asymptotics of the field strengths, it is useful to note
\be
t'_*&=&-r+{\cal O}(1)\quad,\quad {\partial t'_*\over\partial t}={1\over 1+v\sin(\omega t'_*-\phi)},\label{4s2}\\
{\partial t'_*\over\partial x^1}&=& {-\cos\phi\over 1+v\sin(\omega t'_*-\phi)}\quad,\quad 
{\partial t'_*\over\partial x^2}= {-\sin\phi\over 1+v\sin(\omega t'_*-\phi)},\nonumber
\ee
in the $r\to\infty$ limit. These quantities appear in the numerators of the field strength expressions, and one can see from the
structure of the denominators in the above that
in the limit $v\to 1$, these quantities are highly peaked around the angle $\phi_c$ determined by the condition
\be
\phi_c=\omega t'_*(\phi_c)+{\pi\over 2}.\label{4s3}
\ee
We note that $\omega t'_*(\phi)$ is the azimuthal angle of the particle at the intersection of the past light-cone from $(r,\phi)$.
This condition has a simple geometrical meaning: the light pulse emitted from one moment of the trajectory 
is highly collimated in the direction of the instant velocity, and travels with the speed of light. 

Below, we will see that
the leading $\gamma$ contribution is indeed confined in the small angular range $\delta\phi\sim {1\over\gamma^3}$ around $\phi_c$. 
For that purpose, it will be useful to have the formula
\be
{\partial t'_*\over\partial \phi}={\rho\sin(\omega t'_*-\phi)\over 1+v\sin(\omega t'_*-\phi)},\label{4s4}
\ee
so that defining a new convenient angular variable $a$ instead of $\phi$
\be
a\equiv \omega t'_*(\phi)-\phi+{\pi\over 2}\quad,\label{4s5}
\ee
we have a relation
\be
{da\over d\phi} = -{1\over 1+v\sin(\omega t'_*-\phi)}=-{1\over 1-v\cos a}\,\label{4s6}
\ee
which can be integrated as
\be
a-v\sin a = -(\phi-\phi_c)\quad,\label{4s7}
\ee
using the boundary condition (\ref{4s3}). As $\phi$ runs in the range $(0,2\pi)$, $a$ also runs one cycle $(0,2\pi)$ monotonically
with a very steep gradient of order ${\cal O}(\gamma^2)$ around a small region near $a=0$ ($\phi=\phi_c$) that can be seen in (\ref{4s6}). Therefore, the width $\delta a$ of the collimated light pulse near $a=0$ will translate to the width in $\phi$
\be
\delta \phi \sim {\delta a\over\gamma^2},\label{4ss1}
\ee
near the center $\phi=\phi_c$. We will see shortly that $\delta a \sim {1\over\gamma}$ leading to $\delta\phi\sim {1\over\gamma^3}$.

Another important large $r$ expansion is the one for the proper distance,
\be
(t')^2-(\vec x-\vec x')^2=(t')^2-r^2+2r\rho\cos(\omega t'-\phi)-\rho^2\quad,\nonumber
\ee
that appears in the denominators in the field strength expressions. 
Since the $t'$ integration starts from $t'_*$ to $-\infty$, it is more convenient to shift $t'$ integration by
\be
t'\to -t+t'_*\quad,
\ee
upon which $t\in(0,\infty)$, and the proper distance becomes
\be
t^2-2 t'_* t +2r\rho\sin(a-\omega t)-2r\rho\sin a,\label{4s9}
\ee
where we used (\ref{4s5}) and the definition of $t'_*$ (\ref{deft}).
Note that $t_*=-r+{\cal O}(1)$ in $r\to\infty$, and one can consider two parametric regions of the $t$-integral: (1) $t\ll |t_*'|\approx r$
and (2) $t\ge |t_*'|\approx r$.
For (1), one can clearly neglect the  $t^2$ term in (\ref{4s9}) and the proper distance is ${\cal O}(r)$. For (2) the proper distance
becomes of order ${\cal O}(r^2)$, and considering that this enters the denominators of the field strength expressions,
we see that the region (2) of $t$ integral gives us a sub-dominant large $r$ behavior of field strengths.
One concludes that the leading large $r$ value of field strengths arises only from the range (1), and therefore we can
neglect the $t^2$ term in (\ref{4s9}) when computing leading large $r$ asymptotics, and the proper distance can simply be replaced by
\be
2r\left(t+\rho\sin(a-\omega t)-\rho\sin a\right),\label{4s10}
\ee
for this purpose.

It is straightforward to use (\ref{app10}), (\ref{app11}), and (\ref{app12}) to compute the large $r$ asymptotics of the field strengths.
After some algebra, one finds that 
\be
F_{12}=\left(F_{01}\sin\phi-F_{02}\cos\phi\right)\quad,
\ee
at leading order in $r\to\infty$, and
\be
F_{12}&\to& {e\over 2\pi}{v\omega^{1\over 2}\over \sqrt{2r}}{1\over 1-v\cos a}\int_0^\infty dt\,\Bigg\{\nonumber\\
&&{-\cos(a-t)\over\sqrt{t+v\sin(a-t)-v\sin a}}\label{4s11}\\&&+{1\over 2}{v\sin(a-t)\left(\cos(a-t)-\cos a\right)
\over \left(t+v\sin(a-t)-v\sin a\right)^{3\over 2}}\Bigg\},\nonumber
\ee
where we have rescaled $t$ by $\omega t \to t$.
The above expression is a general result for arbitrary velocity $v=\rho\omega$ and the angle $a$ (or equivalently $\phi$).

Now, we need to find the leading $\gamma$ behavior of the total radiated power (\ref{4s1}),
\be
{dE\over dt} = \lim_{r\to\infty} r\int_0^{2\pi}da\,\left|{da\over d\phi}\right|^{-1}|F_{12}|^2,\label{4s12}
\ee 
where we changed the angle integration from $\phi$ to $a$ for convenience.
By careful inspection of the above integral, 
it can be shown that the leading $\gamma$ behavior of ${\cal O}(\gamma^2)$ arises from the narrow range of $(\delta t,\delta a)\sim {1 \over\gamma}$ around $(t,a)=0$, and one can for example expand the denominator as
\be
t+v\sin(a-t)-v\sin a \approx {1\over 2}\left({1\over\gamma^2}+a^2\right)t -{1\over 2} a t^2 +{1\over 6}t^3,\nonumber
\ee
up to the relevant order. One also expands
\be
\left|{da\over d\phi}\right|={1\over 1-v\cos a}\approx {2\over {1\over\gamma^2}+a^2},\label{4s14}
\ee 
as well as the numerator, and by rescaling the integration variables $(t,a)\to {1\over\gamma}(t,a)$, 
the leading $\gamma$-piece can easily be shown to reduce to
\be
{d E\over dt} = {e^2\over (2\pi)^2}\gamma^2 \omega \int_{-\infty}^{\infty} da\, {1\over 1+ a^2} |f(a)|^2,
\ee
where
\be
f(a)=\partial_a\left(\int_0^\infty dt\,{(a-t)\over\sqrt{{1\over 2}(1+a^2)t-{1\over 2} a t^2 +{1\over 6} t^3} }\right).
\ee

The integral 
\be
\int_{-\infty}^{\infty} da\, {1\over 1+ a^2} |f(a)|^2 
\ee
is $(2\pi) {2\over \sqrt{3}}$ up to 4 digits numerically. Assuming this numerical result to be exact, we obtain for the 
total power radiated 
\be
{dE\over dt} = {e^2 \over \sqrt{3}\pi}\omega\gamma^2,\label{4s15}
\ee
which agrees precisely with the rate of work done locally by the ``self-force" (\ref{work}) derived in section \ref{sec21}.

\begin{figure}[t]
	\centering
\includegraphics[width=8cm]{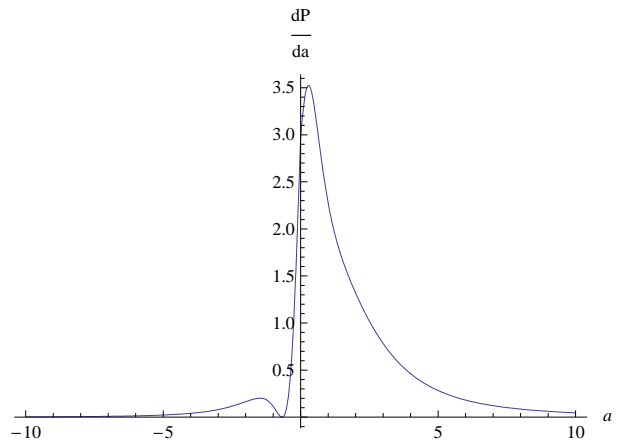}
		\caption{ The angular distribution of radiated power in the leading $\gamma$ approximation. See (\ref{4s7}) for the relation between $a$ (more precisely ${a\over \gamma}$) and $\phi$. \label{fig1}}
\end{figure}

For completeness, in Fig.\ref{fig1}, we plot the angular distribution of the radiation power, ${dP\over da}={1\over 1+ a^2} |f(a)|^2$, as a function of $a$ (recall that the true $a$ is given by $a\over\gamma$).
The plot can be translated to the angular distribution in $\phi$ using the relation (\ref{4s7}).

\section{Summary }

   We have shown that, unlike in the 3+1 dimensional space-time, 
  the  self-force can be defined to be nonzero in odd-dimensional  space-times,
  as the retarded propagators in these cases contain a theta-function
  part or inside-the-light-cone contributions. 

Furthermore, in  the ultrarelativistic case it can be put in local form, depending
only on the instantaneous derivatives of the particle motion.
We have explicitly derived such expressions for the self-force, in 2+1 and 4+1 dimensions.
In the former case we have also calculated the radiation intensity at large distances
and checked that it matches the work done by the self-force numerically. We expect the same 
in 4+1 dimensions with minimal subtraction.

We think it is
perhaps the first instance of an entirely local and consistent derivation of the radiation braking force. 
Although the MSTQW approach   \cite{Mino:1996nk,Quinn:1996am} is
15 years old and it had inspired our work, we are not aware of its explicit tests
in the  gravitational setting either.  Needless to say, much more work is needed
in order to find the exact applicability domain of the ``self-force" approach, in flat and curved space-times.
In~\cite{Shuryak:2011ge} we have suggested that the MSTQW equation when adapted to thermal AdS
may be of relevance to jet quenching in ultrarelativistic collisions such as RHIC and LHC.

\vskip .25cm {\bf Acknowledgments.} \vskip .2cm This work was
supported in parts by the US-DOE grant DE-FG-88ER40388. 


\section{Appendix: Alternative regularization of the self-force in 2+1 dimensions}

In this appendix, we present another way of regularizing the self-force in 2+1 dimensions, by averaging
the Lorentz force
\be
f^\mu_{3D}=e F^{\mu\nu}\left(x(\tau)\right) \dot{x}_\nu(\tau) \quad,\label{app1}
\ee
around a small circle of radius $r$, and taking $r\to 0$ limit after renormalizing the diverging mass.
We will restrict ourselves to the case of ultra-relativistic circular motion for simplicity, 
\be
x^\mu(\tau)=\left(\gamma\tau, \rho\,{\rm cos}(\gamma\omega\tau), \rho\,{\rm sin}(\gamma\omega\tau)\right)\quad,
\label{app5}
\ee 
where $v=\rho\,\omega\approx 1$,
and will find
that the leading $\gamma$ result of the finite self-force agrees with the one in section \ref{sec21}, which
is a confirming check for our results.

Starting from the expression of retarded gauge potential (\ref{ret}), 
\be
A^\mu(x)
=e\int d\tau'\,\Delta_R\left(x-x(\tau')\right)\dot{x}^\mu(\tau')\quad,\label{app2}
\ee
with
\be
\Delta_R(x)={\theta(x^0)\over 2\pi}{\theta(x^2)\over\sqrt{x^2}}\quad,\quad x^2\equiv x^\mu x_\mu\quad,\label{app3}
\ee
we need to compute field strengths at radius $r$ from the position of the charge at $\tau=0$ in
the two-dimensional spatial plane $\vec x=(x^1,x^2)$. 
Let us show the computation $F_{01}=\partial_0 A_1-\partial_1 A_0$ in some detail to set-up notations and procedures,
and we will present other components of $F_{\mu\nu}$ at the end without much details.
The expression (\ref{app2}) gives us
\be
A_0(t,\vec x)={e\over 2\pi}\int^{t'_*(t,\vec x)}_{-\Lambda}\, {dt'\over\sqrt{(t-t')^2-(\vec x-\vec x')^2}}\quad,\label{app6}
\ee
where $\vec x'\equiv \vec x(t')$ is the past trajectory (\ref{app5}) parameterized here in terms of regular time $t'=\gamma \tau'$, and the integration starts from $t'_*(t,\vec x)<0$ which is determined by the intersection between the past light-cone from
the position $(t,\vec x)$ and the particle trajectory,
\be
(t-t'_*)^2=\left(\vec x-\vec x'(t'_*)\right)^2\quad.\label{app7}
\ee
The expression itself is infrared divergent and we introduced a cutoff $\Lambda$, but the field strengths which are physical are
completely IR finite as one takes $\Lambda\to\infty$ at the end, as will be clear in the following.
To compute $\partial_1 A_0$, one needs to evaluate the variation of (\ref{app6}) with respect to $x^1\to x^1+\delta x^1$.
The variation will shift both the integration range through $t'_*$ and the integrand. The former variation naively
gives us the contribution which is proportional to the value of the integrand at $t'=t'_*$ which happens to be divergent due to
(\ref{app7}). The variation of the integrand also gives us an integral which is divergent near $t'=t'_*$.
However, since the original (\ref{app6}) is completely finite near $t'=t'_*$ these divergences are  mere artifacts of improper
manipulations, and in fact the two divergences cancel with each other. A better way of handling them is the following.
From the expression of $A_0$ at $\vec x+\delta\vec x$,
\be
A_0(\vec x+\delta\vec x)={e\over 2\pi}\int^{t'_*+{\partial t'_*\over\partial\vec x}\cdot\delta \vec x}_{-\Lambda}\, {dt'\over\sqrt{(t-t')^2-(\vec x+\delta\vec x-\vec x')^2}},\nonumber\\
\label{app8}
\ee
one shifts the $t'$ integral by $t'\to t'+{\partial t'_*\over\partial\vec x}\cdot\delta\vec x$ so that
the new $t'$ integral starts at the same $t'_*$ while the integrand gets additional contributions
\be
&& A_0(\vec x+\delta\vec x)=\label{app9}\\&&
{e\over 2\pi}\int^{t'_*}_{-\Lambda}\, {dt'\over\sqrt{(t-t'-{\partial t'_*\over\partial\vec x}\cdot\delta\vec x)^2-(\vec x+\delta\vec x-\vec x'-{\partial\vec x'\over\partial t'} {\partial t_*'\over\partial\vec x}\cdot\delta\vec x)^2}},\nonumber
\ee
which is free from the divergence near $t'=t'_*$. 
There is also a shift in the IR cutoff 
\be
\Lambda\to\Lambda-{\partial t'_*\over\partial\vec x}\cdot\delta\vec x\quad,\ee
but it can be easily shown that it becomes irrelevant in the final results, and we omit it in (\ref{app9}).

Taking difference between (\ref{app9}) and (\ref{app6}) to first order in $\delta\vec x$, one readily computes 
$\vec\partial A_0$ as (after putting $t=0$)
\be
&& \vec\partial A_0 =\label{app10}\\
&& {-e\over 2\pi}\int^{t'_*}_{-\infty}dt'\,{{\partial t'_*\over\partial\vec x} t'-(\vec x-\vec x')
+(\vec x-\vec x')\cdot{\partial \vec x'\over\partial t'} {\partial t'_*\over\partial\vec x} \over
\left((t')^2 -(\vec x-\vec x')^2\right)^{3\over 2}},\nonumber
\ee
where we removed $\Lambda$ to infinity as the final integral is finite.

By similar steps, one obtains
\be
&&\partial_0 \vec A={-e\over 2\pi}
\int^{t'_*}_{-\infty}dt'\,\Bigg({{\partial^2\vec x'\over\partial t'^2}{\partial t'_*\over\partial t}\over \sqrt{(t')^2-(\vec x-\vec x')^2}}\label{app11}\\
&&+{{\partial\vec x'\over\partial t'}\left((1-{\partial t'_*\over\partial t})t' -(\vec x-\vec x')\cdot{\partial\vec x'\over\partial t'}{\partial t'_*\over\partial t}\right)\over \left((t')^2 -(\vec x-\vec x')^2\right)^{3\over 2}}\Bigg).\nonumber
\ee
The above (\ref{app10}) and (\ref{app11}) give us the field strength $F_{0i}=\partial_0 A_i-\partial_i A_0$, $i=1,2$.
The expressions for ${\partial t'_*\over \partial t}$ and $\partial t'_*\over\partial \vec x$ that appear in the above
can be easily obtained from the relation (\ref{app7}).
Finally, $F_{12}=\partial_1 A_2-\partial_2 A_1$ is written as
\be
&&F_{12}={e\over 2\pi}
\int^{t'_*}_{-\infty}dt'\,\Bigg({\epsilon_{ij}{\partial^2\vec x'^i\over\partial t'^2}{\partial t'_*\over\partial x^j}\over \sqrt{(t')^2-(\vec x-\vec x')^2}}\label{app12}\\
&&-{\epsilon_{ij}{\partial\vec x'^i\over\partial t'}\left({\partial t'_*\over\partial x^j}t' -(x^j-x'^j) +(\vec x-\vec x')\cdot{\partial\vec x'\over\partial t'}{\partial t'_*\over\partial x^j}\right)\over \left((t')^2 -(\vec x-\vec x')^2\right)^{3\over 2}}\Bigg),\nonumber
\ee
where $\epsilon_{12}=-\epsilon_{21}=+1$. 
We will apply the above formulae to our case of circular motion (\ref{app5}).

We then consider a small circle of radius $r$ from the position of the charge at $t=0$, that is, $(\rho,0)$.
We let the azimuthal angle of the circle be $\phi$, so that a point on the circle has the coordinate $\vec x=(\rho+r\cos\phi,r\sin\phi)$. We will compute the Lorentz force on the points in the circle and take an average over $\phi$
before taking the limit $r\to 0$.
Noting the 3-velocity ${dx^\mu\over d\tau}\equiv u^\mu =\gamma (1,0,v)$ at $t=0$ (recall $v=\rho\omega$),
let's first look at the transverse component of the self-force,
\be f^1=e\gamma \left(F_{01}-v F_{12}\right)\quad.
\ee
Near $r\to 0$, the useful expansions of some quantities are
\be
t'_*&=& \gamma^2 xr + {\cal O}(r^2)\,\,,\,\,x\equiv -({v\sin\phi+\sqrt{1-v^2\cos^2\phi}})\nonumber,\\
{\partial t'_*\over \partial x^1}&=&{\cos\phi\over {x}+v\sin\phi}\quad,\quad
{\partial t'_*\over \partial x^2}={\sin\phi-\gamma^2 v x \over {x}+v\sin\phi},\nonumber\\
{\partial t'_*\over \partial t}&=&{\gamma^2 x\over {x}+v\sin\phi}\quad,\label{app13}
\ee
and from these, one can derive that the $\phi$-averaged transverse force has a divergence near $r\to 0$ from the term
\be
f^1&\sim& {e^2\gamma^2 v\omega\over 2\pi}\int^{\gamma^2 x r} dt'{1\over\sqrt{(t'-\gamma^2 rx)(t'-\gamma^2 r x')}}\nonumber\\
&\sim&{e^2\gamma^2 v\omega \over 2 \pi} \log\left(1\over r\right)=-{e^2\over 2\pi}\log\left(1\over r\right)\ddot{x}^1,
\label{app14}
\ee
where $x'=-({v\sin\phi-\sqrt{1-v^2\cos^2\phi}})>0$. In deriving this, we have used an important expansion
\be
(t')^2-(\vec x-\vec x')^2 \approx {1\over\gamma^2}(t'-\gamma^2 rx)(t'-\gamma^2 r x'),\label{app15}
\ee
near $r\to 0$ , ${t'\over r}\sim {\cal O}(1)$ limit, which gives one factor of $\gamma$ in (\ref{app14}).
This divergence is precisely of the same character of the mass renormalization we encountered in section \ref{sec21},
and one can absorb it by 
\be
m_{ren}=m_{bare}+{e^2\over 2\pi}\log\left(1\over r\right)\quad.
\ee

We are more interested in the leading $\gamma$ behavior of longitudinal self-force,
\be
f^2=e\gamma F_{02}\quad,\label{app15}
\ee
and using (\ref{app13}), averaging over $\phi$ and taking $r\to0$ limit, one arrives at a finite expression,
\be
f^2&=&-{e^2\gamma^3\over 2 \pi}\int^\infty_0 dt'\, \Bigg({v\omega\sin(\omega t')\over\sqrt{
(t')^2-4\rho^2\sin^2\left(\omega t'\over 2\right)}}\nonumber\\
&+& {\rho\left(1-v^2\cos(\omega t')\right)\left(\sin(\omega t')-\omega t'\right)\over
\left((t')^2-4\rho^2\sin^2\left(\omega t'\over 2\right)\right)^{3\over 2}}\Bigg),\label{app16}
\ee
where we changed the variable $t'\to-t'$ in the integration. The last step is to find a leading $\gamma$ behavior of (\ref{app16}). Rescaling $\omega t'\to t$, we have
\be
f^2&=&-{e^2\gamma^3 v\omega\over 2\pi} \int^\infty_0 dt\, \Bigg({\sin t\over\sqrt{
t^2-4v^2\sin^2\left(t\over 2\right)}}\nonumber\\
&+& {\left(1-v^2\cos t\right)\left(\sin t - t\right)\over
\left(t^2-4v^2\sin^2\left(t\over 2\right)\right)^{3\over 2}}\Bigg).\label{app17}
\ee
To study the large $\gamma$ behavior of the integral in (\ref{app17}) replacing $v^2=1-{1\over\gamma^2}$, one writes the integral
after some manipulations
\be
&&\int_0^\infty dt\,\partial_t\left({(1-\cos t)\over \sqrt{t^2-4\sin^2\left(t\over 2\right)}}\right)\\
&& +{1\over\gamma^2}\int_0^\infty dt\,{4\sin t\sin^2\left(t\over 2\right)+(\sin t-t)\cos t\over \left(t^2-4\sin^2\left(t\over 2\right)+{4\over\gamma^2}\sin^2\left(t\over 2\right)\right)^{3\over 2}},\nonumber
\ee
neglecting additional ${\cal O}(\gamma^{-2})$ contributions. The first integral is completely localized at $t=0$ giving a value of $-\sqrt{3}$. In the second integral, one can show that a leading ${\cal O}(1)$ result comes from a range of small $t\sim {\cal O}\left(1\over\gamma\right)$, and  
one has an approximation in the denominator,
\be
t^2-4\sin^2\left(t\over 2\right)+{4\over\gamma^2}\sin^2\left(t\over 2\right)\approx {t^2\over\gamma^2}+{1\over 12}t^4 ,\label{app18}
\ee 
where higher order terms can be shown to be irrelevant in the leading $\gamma$ contributions.
The $t^4$ term in (\ref{app18}) effectively cutoffs the $t$ integral to be $t\le {1\over \gamma}$, which we have seen
before in section \ref{sec21} (recall the proper time is $\tau={t\over\gamma}$).
The small $t$ expansion for the numerator up to relevant order is
\be
4\sin t\sin^2\left(t\over 2\right)+(\sin t-t)\cos t\approx{5\over 6}t^3 ,\label{app20}
\ee
giving us the second integral
\be
{1\over\gamma^2}\int_0^\infty dt\,{{5\over 6}t^3\over\left({t^2\over\gamma^2}+{1\over 12}t^4\right)^{3\over 2}} ={5\over 3}\sqrt{3}.\label{app21}
\ee
In total, the integral in (\ref{app17}) gives us $(-1+{5\over 3})\sqrt{3}={2\over\sqrt{3}}+{\cal O}(\gamma^{-2})$, so that the leading $\gamma$ result of $f^2$ is
\be
f^2\sim -{e^2\omega\gamma^3 \over \sqrt{3}\pi}\sim -{e^2\omega \gamma^2\over\sqrt{3}\pi} \dot{x}^2,\label{app19}
\ee
which agrees precisely with (\ref{ressec2}) in section \ref{sec21}.

\end{document}